\documentclass[11pt,a4paper]{article}

\usepackage{graphicx}
\DeclareGraphicsExtensions{.eps}

\begin{document}

\author{B. E. Sauer, D. M. Kara, J. J. Hudson, M. R. Tarbutt, E. A. Hinds\\
\emph{Centre for Cold Matter, Blackett Laboratory,}\\ 
\emph{Imperial College London, Prince Consort Road,}\\
\emph{London SW7 2AZ. United Kingdom.}}
\title{A robust floating nanoammeter}
\maketitle

\begin{abstract}
A circuit capable of measuring nanoampere currents while floating at voltages up to at least 25kV is described. 
The circuit
relays its output to ground potential via an optical fiber. We particularly emphasize the design and construction
techniques which allow robust operation in the presence of high voltage spikes and discharges.
\end{abstract}

It is occasionally necessary to measure nanoamp or smaller currents flowing in a part of a circuit at high voltage. 
It is often easiest to
do so by making a measurement of the ground return current of the high voltage supply \cite{HG}, 
but occasionally this is 
inconvenient. Our particular application stems from an experiment to measure the permanant electric dipole moment of
the electron \cite{JJH} \cite{ICAP}, in which voltages of up to $\pm 12$kV are applied to electric field plates in order to 
polarize a beam of YbF molecules. A leakage current larger than about 10nA between the field plates is a  
source of systematic error in the experiment. It is desirable to measure directly this plate leakage current, which 
is orders of magnitude smaller than the total current drawn from the the high voltage supply. Most of this current 
does not contribute to the possible systematic error, but flows to ground though various 
divider chains and bleed resistors which are part of the high voltage circuit.

The obvious solution is to float the sensitive measurement circuitry at high voltage and communicate its output 
optically to a
recording instrument at ground potential \cite{ANA}. This can be done with varying levels of sophistication. 
Some methods which
have been implemented include a floating analog-to-digital converter and UART for parallel to serial conversion with
transmission via an optical fiber link \cite{PEA}, voltage-to-frequency conversion followed by optical transmission
\cite{LI}, and direct voltage transmission through a commercial optically isolated buffer amplifier \cite{JAD}. This last
method is limited to a few kV. In contrast, a glass or plastic optical fiber link can withstand very large
potential differences.

We have built a number of different floating ammeters and found there are two issues which tend to limit performance:
the power supply for the floating circuitry and the failure of semiconductor components after high voltage discharges.
The first is the easier to solve. For an instrument which is used intermittently, battery power is a simple solution. We
have had success using both disposable alkaline and rechargable lead acid batteries. In the electron electric dipole 
moment experiment, however, the flattening of the supply
batteries is not acceptable because the ammeter is required to run continuously for 
days. Instead it is powered by a solar cell mounted a few cm from a bank of high power
LEDs. 
This is remarkably inefficient, with some 30W of electrical power required to power the red LED array \cite{led}, 
while the 
solar cell delivers less than 0.5W to the ammeter circuit. The solar cell has higher
conversion efficiency for green rather then red light but this is more than compensated by the higher output 
power available from red LEDs.
We find this method of powering the ammeter perfectly
reliable. The solar cell power does entail the slight complication of single polarity supply, but Fig. 1 shows one way 
to bias the input circuitry in order to record bipolar currents.
Isolated DC to DC converters are available commercially and would be preferable as power sources for 
circuits which only need
to float at a few kV. Other methods to deliver power to the high voltage instrument are isolation transformers or
small DC motors coupled via a dielectric shaft and run as a motor-generator pair. We have not implemented these 
latter solutions.

\begin{figure}[bt]
\begin{center}
\includegraphics[width=12cm]{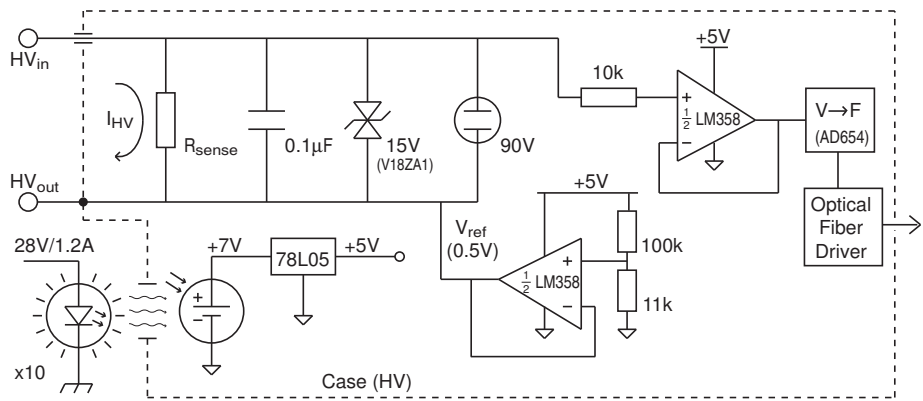}
\caption{The nanoammeter circuit, with some of the details of the input circuitry shown explicitly.
Not shown is the standard circuitry associated with the voltage to frequency converter and optical fiber driver. 
A number of capacitors have been omitted for clarity. The expected path for the 
high voltage current $I_\mathit{HV}$ is shown explicitly. The Faraday case (see text) is tied to the terminal labelled
`HV$_{out}$'. The bias scheme for the sense
resistor places the high voltage common 0.5V below the case, as is appropriate for the AD654 voltage-to-frequency 
converter. The ten high power LEDs which drive the solar cell are wired in series.} \label{frontend}
\end{center}
\end{figure}

Failure of the ammeter due to high voltage events has proved to be a more serious issue. We have designed and built 
several ammeters which have broken down in spectacular fashion. Figure 1 shows the front end of an ammeter that was
designed to solve this problem. It was extensively tested with a spark gap load, which triggered between 5kV and 10kV \cite{SG}. 
Ammeters of this design have not failed in over a year of actual operation in our laboratory. 
Note that in this circuit the op-amp is configured as a simple voltage buffer rather than as a transresistance amplifier.
This means that current surges should flow through the protective elements rather than the op-amp. The 10k$\Omega$
resistor in series with the op-amp input serves solely to isolate the op-amp from the HV current path. The sensitivity of 
the ammeter is determined by the current
sensing resistor, $R_\mathit{sense}$, which is typically between 0.1M$\Omega$ and 10M$\Omega$. 
The varistor (Littlefuse V18ZA1) has a very large resistance
below 10V, dropping to about 1M$\Omega$ at 15V and 2k$\Omega$ at 20V. The gas discharge element provides a circuit path
for very large current spikes.

For the optical link, our device uses multimode plastic fiber and an Avago HFBR 1521/2521 transmitter/receiver pair. The optical
transmitter dominates the power consumption in the circuit, thus no special effort was made to save power elsewhere in the circuit. 
To conserve power
the optical transmitter drive was reduced to a level that was just able to drive a 0.5m multimode fiber. 
This length was adequate to reach an optical 
repeater operating at high power near ground potential. The
venerable LM358 was found to be adequate as an amplifier, though there is no reason that any of the multitude of modern micropower
amplifiers would not prove to be equally effective. Figure 1 shows an Analog Devices AD654 as the voltage-to-frequency converter. 
This particular device has an input range of 0 to 1V, thus one side of the sense resistor is biased to 0.5V in order to be
sensitive to bipolar currents. We have also constructed ammeters using components from other manufacturers, for which the
details of the biasing scheme were different. For the circuit in fig. 1 with a sense resistor of 1M$\Omega$ and a full scale
V-F output of 200kHz, the sensitivity of the ammeter is 200Hz/nA.

Our first ammeter designs suffered from leakage from the ammeter components themselves. In our current design, all of the 
floating components are placed in a diecast aluminium box which is attached to the high voltage output and acts as a 
Faraday cage. 
In turn, this box sits on nylon standoffs inside a larger polycarbonate box. The high voltage connections are 
made through the top of the outer box, which is hermetically sealed to allow the possiblity of filling the instrument with
transformer oil or SF$_6$ to prevent breakdown. At voltages up to $\pm 25$kV this has not proved necessary.
The ammeter circuit board itself is coated with anti-corona lacquer.

Figure 2 illustrates the performance of a pair of floating ammeters. It shows the charging currents which flow after the
polarity of a pair of electric field plates is switched. The charging time constants are determined by the combined 
capacitance of the
plates and their connecting cables and a series resistor which limits the peak current. Note the residual leakage current is
very small, less than 2nA. This leakage data is recorded continuously as part of the experiment to measure the electron
electric dipole moment.

\begin{figure}[bt]
\begin{center}
\includegraphics[width=8cm]{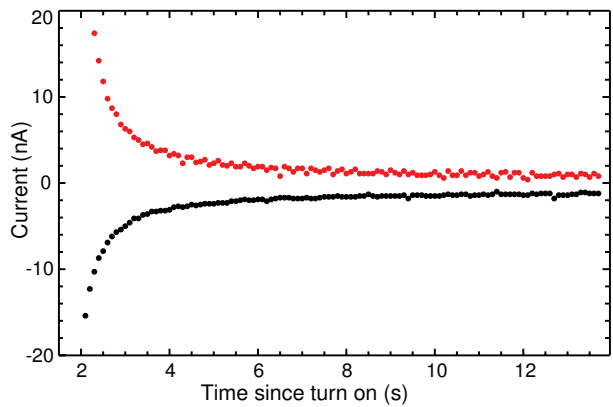}
\caption{Currents recorded as a pair of electrodes are charged to $\pm7$kV. The pulse output from the floating nanoammeters
was recorded in 50ms bins every 100ms. The first 2 seconds of the charging currents (some of which is out of range)
are not shown.} \label{fig2}
\end{center}
\end{figure}

\emph{Detailed circuit diagrams are available from the authors on request.}


\begin{thebibliography}{30}
\bibitem{SG}A simple adjustable spark gap can be made with sharpened machine screws and a wooden block.
\end{thebibliography}
\end{document}